\documentclass[useAMS,usenatbib]{mn2e}

\usepackage{rotating}
\usepackage{graphicx}
\usepackage{txfonts}
\usepackage{lscape}
\usepackage{natbib}
\bibpunct{(}{)}{;}{a}{}{,} 

\title[Clumpy galaxies at z$\sim$0.6]{Clumpy galaxies at z$\sim$0.6:
  kinematics, stability, and comparison with analogs at other
  redshifts} \author[M. Puech]{M. Puech$^{1}$\thanks{E-mail:
    mathieu.puech@obspm.fr}\\ $^{1}$GEPI, Observatoire de Paris, CNRS,
  University Paris Diderot; 5 Place Jules Janssen, 92195 Meudon Cedex,
  France}
\begin{document}

\date{Accepted ...}

\pagerange{\pageref{firstpage}--\pageref{lastpage}} \pubyear{2002}

\maketitle

\label{firstpage}

\begin{abstract}
Distant clumpy galaxies are thought to be Jeans-unstable disks, and an
important channel for the formation of local galaxies, as suggested by
recent spatially-resolved kinematic observations of z$\sim$2 galaxies.
I study the kinematics of clumpy galaxies at z$\sim$0.6, and compare
their properties with those of counterparts at higher and lower
redshifts. I selected a sample of 11 clumpy galaxies at z$\sim$0.6
from the representative sample of emission line, intermediate-mass
galaxies IMAGES. Selection was based on rest-frame UV morphology from
HST/ACS images, mimicking the selection criteria commonly used at
higher redshifts. Their spatially-resolved kinematics were derived in
the frame of the IMAGES survey, using the VLT/FLAMES-GIRAFFE
multi-integral field spectrograph. For those showing large-scale
rotation, I derived the Toomre $Q$ parameter, which characterizes the
stability of their gaseous and stellar phases. I find that the
fraction of UV-selected clumpy galaxies at z$\sim$0.6 is 20$\pm$12\%.
Roughly half of them (45$\pm$30\%) have complex kinematics
inconsistent with Jeans-unstable disks, while those in the remaining
half (55$\pm$30\%) show large-scale rotations. The latter reveal a
stable gaseous phase, but the contribution of their stellar phase
makes them globally unstable to clump formation. Clumpy galaxies
appear to be less unstable at z$\sim$0.6 than at z$\sim$2, which could
explain why the UV clumps tend to vanish in rest-frame optical images
of z$\sim$0.6 clumpy galaxies, conversely to z$\sim$2 clumpy galaxies,
in which the stellar phase can substantially fragment. This suggests
that the former correspond to patchy star-formation regions
superimposed on a smoother mass distribution. A possible and
widespread scenario for driving clump formation relies on
instabilities by cold streams penetrating the dark matter halos where
clumpy galaxies inhabit. While such a gas accretion process is
predicted to be significant in massive, z$\sim$2 haloes, it is also
predicted to be strongly suppressed in similar, z$\sim$0.6 haloes,
which could explain why lowest-z clumpy galaxies appear to be driven
by a different mechanism. Instead, I found that interactions are
probably the dominant driver leading to the formation of clumpy
galaxies at z$<$1. I argue that the nature of z$>$1 clumpy galaxies
remains more uncertain. While cold flows could be an important driver
at z$\sim$2, I also argue that the \emph{observed and cumulative}
merger fraction between z=2 and z=3 is large enough so that every
z$\sim$2 galaxy might be the result of a merger that occurred within
their past 1 Gyr. I conclude that it is premature to rule out mergers
as a universal driver for galaxy evolution from z$\sim$2 down to z=0.
\end{abstract}


\section{Introduction}
How galaxies formed, evolved, and built-up the local Hubble sequence
is still an open and highly debated issue. For instance, there is now
evidence for a significant evolution of the stellar mass density at
z$<$1. Even between redshifts as low as z$\sim$0.6 and z=0, the
stellar mass density appears to be evolving by 0.2 dex (i.e.,
$\sim$40\%, see, e.g., \citealt{PerezGonzalez08}). Such an evolution
in stellar mass can also be seen in the evolution of the Tully-Fisher
relation, which appears to be evolving by a factor two in stellar mass
over the same redshift range \citep{puech08,puech09b}. Such an
increase in stellar mass needs to be fed by fresh gas. The absence of
evolution in zero point of the \emph{baryonic} Tully-Fisher relation
between z$\sim$0.6 and z=0 \citep{puech09b} suggests that most of this
gas was already gravitationally bound to galaxies at z$\sim$0.6. What
is the mechanism driving the conversion of this gas into stars? In
spite of the impressive progress accomplished over the past years,
mainly through spatially-resolved kinematics of distant galaxies, this
issue still remains without a clear answer.

At z$\sim$0.6 (i.e., 6 Gyr ago), \cite{yang08} gathered a
representative sample of 63 galaxies with $M_{stellar} \geq 1.5\times
10^{10} M_\odot$ and spatially-resolved kinematics derived from
FLAMES/GIRAFFE observations at the VLT. They found that 40\% of
z$\sim$0.6 intermediate-mass galaxies (i.e., the progenitors of local
spirals), have chaotic velocity fields inconsistent with expectations
from pure rotating disks. Subsequent analyzes suggest that most of
them are likely associated with major mergers \citep{hammer09b}.
Mergers are also found to be a good driver for the large scatter seen
in the distant Tully-Fisher relation \citep{puech09b,covington10}.
Actually, the remarkable co-evolution of the morphology, kinematics,
star formation density, metal density, and stellar mass density, all
could find a common and natural explanation in the frame of the spiral
rebuilding scenario, according to which 50 to 75\% of present-day
spiral disks were rebuilt after a major merger since z=1, as proposed
by \cite{hammer05}. To this respect, the Milky Way appears to be
exceptional, with a remarkable quite past history compared to other
local spiral galaxies \citep{hammer07}.

There is now a growing body of evidence suggesting that disk
rebuilding indeed took place at z$\leq$1, which makes it as a viable
driver for galaxy evolution at these epochs. On the theoretical side,
numerical simulations showed how gas can be expelled in tidal tails
during such mergers, and how it can be subsequently re-accreted
\citep{barnes02}. This re-accreted gas is expected to cool down and
form new stars, re-building a new disk around a spheroidal remnant
\citep{springel05, robertson06}, which might correspond to
morphologies as late as Sb galaxies \citep{lotz08,hopkins09c}. Recent
theoretical developments have shed light on the underlying process,
which appears to be purely gravitational \citep{hopkins09}. The
requirement for a major merger to rebuild a new disk depends mainly on
the gas fraction during the final coalescence, which needs to be at
least 50\% \citep{robertson06}. Cosmological simulations are now also
producing such re-processed disks at z$<$1, although the role of
cosmological gas accretion is not totally understood, but taking this
process into account could result in a lower gas fraction threshold
for rebuilding a new disk \citep{governato08}. On the observational
side, first examples of rebuilt disks were recently detected at
z$\sim$0.6 \citep{puech09,hammer09}, and the auto-consistency of the
disk rebuilding process starts being investigated, both theoretically
\citep{hopkins09b,stewart09,hopkins09c} and observationally
\citep{hammer09b,kannappan09,bundy09,huertas10}.

If the spiral rebuilding scenario appears to achieve encouraging
successes in describing galaxy evolution at z$<$1, some points still
need to be investigated. In particular, the impact of the expected
numerous \emph{minor} mergers on the survival of thin disks is still
debated \citep{toth92,hopkins08,purcell09,moster09b}. Furthermore,
Luminous InfraRed Galaxies (LIRGs) account for about 80\% of the star
formation density reported at z$\leq$1 \citep{hammer05}, but their
morphologies reveal that only half of z$>$0.5 LIRGs are compatible
with mergers, while those in the other half appear to be spiral
\citep{melbourne05}. \cite{marcillac06} showed that the large number
density of LIRGs at these epochs suggests that they could experience
between two and four star formation bursts until z=0, with typical
timescales of $\sim$0.1 Gyr, which is not consistent with a simply
continuous star formation history. They concluded that minor mergers,
tidal interactions, or gas accretion remain plausible triggering
mechanisms in distant LIRGs harboring a spiral morphology.
Interestingly, 75\% of local LIRGs are barred, which could play a role
in regulating star formation in such objects \citep{wang06}.

At higher redshifts (i.e., z$>$1), the co-moving density of galaxy
appears to be dominated by clumpy irregular galaxies
\citep{elmegreen07}. Interest for such objects dates back to
\cite{cowie95}, who first noticed the unusual aspects of some high-z
galaxies, dubbed as ``chain galaxies'' and described as ``linearly
organized giant star-forming regions''. The large occurrence of blue
star-forming knots in less edge-on objects was also later recognized
as a general and intriguing feature of distant galaxies, probably
linked to an early phase in the formation of local spiral galaxies
\citep{cowie95,vandenbergh96}, and were latter referred to as ``clump
clusters'' by \cite{elmegreen04}. The improved spatial resolution
provided by the HST/ACS re-invigorated the interest for these objects,
which were all suggested to be different incarnations of the same
underlying population viewed along different inclination angles
\citep{oneil00,elmegreen04b}. Both kind of objects are therefore often
referred to as ``clumpy galaxies''. They are found to be typically
made of 5-10 kpc-sized clumps with stellar masses $\sim 10^{7-9}$
M$_\odot$ (i.e., $\sim$100 times more massive than the largest star
complexes in present-day spiral galaxies), and they typically account
for one third of the total galaxy emission \citep{elmegreen05}. Such
clumps are thought to be linked to the formation of disks, as
suggested by the increase of the inter-clump surface density, and the
decrease of the mass surface density contrast between the clumps and
the inter-clump regions, when going from clumpy galaxies with no
evident inter-clump emission to clumpy galaxies with faint red disks,
and spiral galaxies \citep{elmegreen09b}.

What is driving the formation of these clumps has been the subject of
many attentions during the past decade. Numerical simulations
suggested that clumps might originate from the local gravitational
instability of very gas-rich disks of young galaxies
\citep{noguchi98,immeli04}. Due to their large masses, the clumps
would experience strong dynamical friction and spiral towards the
galaxy center within a few Gyr, which might lead to the formation of a
bulge \citep{noguchi99,elmegreen08,elmegreen09}, as well as a thick
stellar disk through strong stellar scattering \citep{bournaud09b}.
Simulated clumps are found to show properties similar to observations
\citep{immeli04b,bournaud07}. The lifetime of these clumps is so short
and the fraction of clumpy galaxies at z$>$1 so high, that making the
clumpy phase a long-term phenomenon requires a continuous and rapid
fresh supply of cold gas in order to feed the disk and regenerate new
clumps \citep{dekel09b}. Theoretical developments indeed suggested
that early galaxy formation is fed by cold streams penetrating through
dark matter halos \citep{dekel09}. These cold streams are expected to
maintain a dense disk that can undergo gravitational fragmentation
into several giant clumps \citep{dekel09b}. This possible link between
high-z clumpy galaxies and the cosmological context was strengthened
both by recent cosmological numerical simulations
\citep{agertz09,ceverino09}, and semi-analytic models
\citep{khochfar08}.

Alternatively, it was proposed that clumps could also result from
on-going mergers or interactions
\citep{taniguchi01,overzier08,dimatteo08}. Discriminating between the
merger and fragmentation scenarii is not straightforward because it
requires high-resolution integral field spectroscopy in high-z
galaxies. Indeed, as stated by \cite{noguchi99}, the most
straightforward and powerful test for discriminating between the two
hypothesis is to examine the kinematics of the clumps: in the merger
scenario, a random orientation of clump spins is expected, while in
the fragmentation scenario, clumps are expected to be coplanar. In
particular, one interesting predictions of these simulations is that
the large-scale rotation in the underlying disk should be preserved
during the fragmentation phase \citep{immeli04,bournaud07}. The
achievement of integral field spectrograph working in the near
infrared (e.g., SINFONI at the VLT, or OSIRIS at Keck), allowed
several teams to gather spatially-resolved kinematic observations of
z$>$1 distant galaxies (see \citealt{forsterschreiber09,law09} and
references therein, as well as
\citealt{wright07,wright08,bournaud08,vanstarkenburg08,epinat09}).
Detection of an underlying rotation was claimed in several z$\sim$2
clumpy galaxies, which has been used as a support to the fragmentation
scenario \citep{genzel08}.

Surprisingly, analysis of integral field spectroscopy observations
suggest that two different scenarii might take place at z$<$1, and at
z$>$1. It therefore becomes necessary to start investigating whether
both scenarii are consistent, and whether or not a transition between
two different galaxy evolution drivers is occurring between these two
epochs. To this aim, I take advantage of the IMAGES survey to study
clumpy galaxies at z=0.6. The goal is to investigate which of the
merger or fragmentation scenario is the most consistent with
z$\sim$0.6 clumpy galaxies, and investigate whether clumpy galaxies at
z$<$1 and at z$>$1 are driven by a common physical process. This paper
is organized as follows: In Sect. 2, I describe how the clumpy
galaxies at z$\sim$0.6 were selected; In Sect. 3, I present their
kinematic and dynamical properties; Sect. 4 discusses the origin of
the clumpiness in distant galaxies; Sect. 5 discusses the results,
while conclusions are drawn in Sect. 6. Throughout the paper, I adopt
$H_0=70$ km/s/Mpc, $\Omega _M=0.3$, and $\Omega _\Lambda=0.7$, and the
$AB$ magnitude system.

\section{Sample selection \& fraction of clumpy galaxies at z$\sim$0.6}
In this paper, I made use of the representative IMAGES sample of 63
emission line galaxies at z$\sim$0.6. The sample is fully described in
\cite{yang08}, and I refer the reader to this paper for details. This
sample was observed by FLAMES/GIRAFFE at the VLT, which allowed us to
derived spatially-resolved kinematics for all these galaxies. The
dynamical state of these galaxies was studied, according to which they
were classified into three classes: Rotating Disks (RDs) are galaxies
showing regular rotation well-aligned along the morphological axis,
Perturbed Rotators (PRs) show a large-scale rotation with a local
perturbation in the velocity dispersion map that cannot be accounted
for by rotation, while galaxies with Complex Kinematics (CKs) do not
show any large-scale rotation, or a strong misalignment between the
dynamical and morphological axes \citep{yang08}. This classification
takes into account the residuals between the observed VF and
$\sigma$-map and those predicted by a rotating-disk model (see
\citealt{yang08}), which makes this classification objective and
reproducible. For reasons of homogeneity, and to benefit from the most
exquisite images from the ACS camera on-board the HST, I limited
myself to galaxies lying in the CDFS-GOODS field. This study is
therefore based on a sample of 32 galaxies. Note that this sample is
still representative of z$\sim$0.6, intermediate-mass galaxies, as
shown by \cite{yang08}.

\begin{figure*}
\centering
\includegraphics[width=19cm]{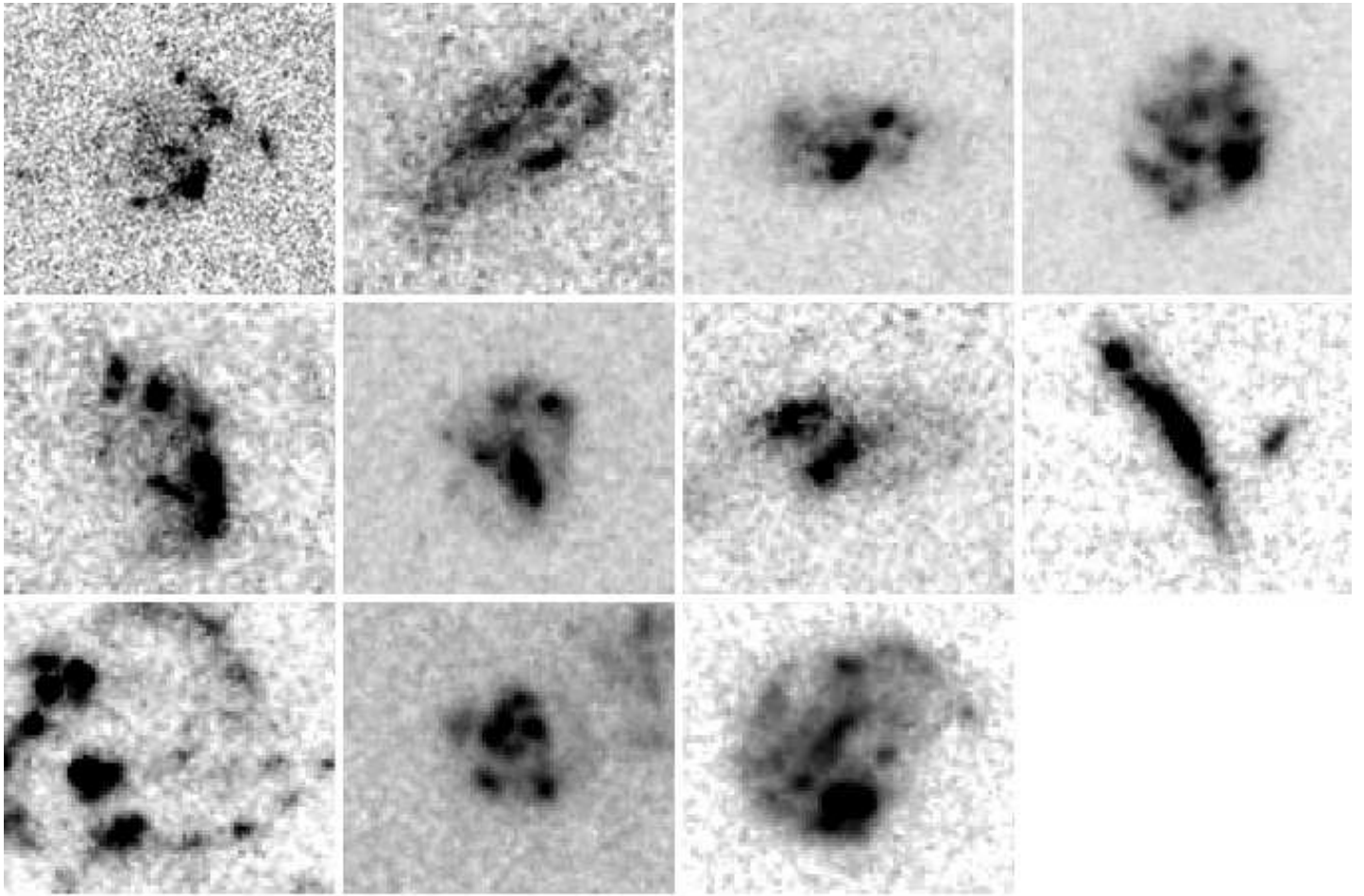}
\caption{Rest-frame UV ($\sim$2700\AA) morphology of the 11 clumpy
  galaxies selected in the IMAGES survey, at z$\sim$0.6. \emph{From
    left to right, and top to bottom:} J033231.58-274121.6,
  J033210.25-274819.5, J033219.61-274831.0, J033233.90-274237.9,
  J033234.04-275009.7, J033239.04-274132.4, J033224.60-274428.1,
  J033227.07-274404.7, J033230.57-274518.2, J033234.12-273953.5,
  J033239.72-275154.7.}
\label{morpho}
\end{figure*}

I selected clumpy galaxies following selection criteria depicted by
\cite{elmegreen05} and \cite{elmegreen07}. Specifically, I required
for a galaxy to be clumpy to fulfil the following requirements:
\begin{description}
\item (i) showing at least three clumps in the rest-frame UV image. To
  check this, I used $B_{435}$-band HST/ACS images, which corresponds
  to $\sim$270 nm rest-frame. Note that \cite{elmegreen05} used the
  observed $i_{775}$ band but at higher redshift (see also
  \citealt{elmegreen07}), which also roughly matches the rest-frame UV
  for most of their clumpy galaxies;

\item (ii) absence of spatial features associated with spiral galaxies
  (e.g., central bulge);

\item (iii) the underlying light profile must be inconsistent with an
  exponential disk \citep{elmegreen05b}. Therefore, I discarded all
  galaxies where a Sersic profile resulted in a good fit (see
  \citealt{neichel08}).
\end{description}
Obviously, such a selection method is rather subjective. In order to
stick to former studies of such galaxies at higher redshift, I
nevertheless chose to keep these selection criteria and ignored
further color and kinematic information from the IMAGES database.
Instead, this information was used \emph{a posteriori} to investigated
whether the candidates selected this way were truly rotating and/or
corresponded to Jeans fragmentation processes. The selection was done
independently by three member of the IMAGES consortium (including the
author), and results compared until a consensus was reached for each
object. I ended up with a sample of 11 candidates, which are shown in
Fig. \ref{morpho}. A visual comparison with Fig. 1 of
\cite{elmegreen05} confirms that all z$\sim$0.6 candidates indeed look
similar to higher-z clumpy galaxies.

Since the IMAGES-CDFS sample is representative of z$\sim$0.6 emission
line galaxies, I can estimate the fraction of emission line clumpy
galaxies at z$\sim$0.6, which is found to be 34$\pm$18\%, with
error-bars due to Poisson fluctuation in the parent sample. This is in
very good agreement with results from the UDF at z$<$1, and a factor
$\sim$2 less than at z$\sim$2-3 \citep{elmegreen07}. If one accounts
for the fact that emission line galaxies represent 60\% of
intermediate mass galaxies at z$\sim$0.6 (\citealt{hammer97}; see also
\citealt{mignoli05}), one finds that clumpy galaxies represent
20$\pm$12\% of intermediate-mass galaxies at this redshift.

\section{Kinematics \& Stability of z$\sim$0.6 clumpy galaxies}

\subsection{Clumpy galaxies with complex kinematics: mergers?}
Spatially-resolved kinematics was obtained, amongst others, for all of
the 11 clumpy galaxies by \cite{yang08}. Amongst them, five were found
to be CKs, four corresponded to PR, and two were classified as RDs
(see Appendix). Rotation requires a significant inclination angle to
be detected, (i.e., a face-on disk would not show any rotation due to
projection effects). I checked that such projection effects do not
affect the relative fraction of CK galaxies: considering the full
sample of 63 IMAGES galaxies from \cite{yang08}, the fraction of
face-on galaxies (i.e., having an inclination angle smaller than 30
degrees), is found to be 10, 17, and 13\%$\pm$8\% amongst RD, PR, and
CK galaxies, respectively. This rules out any significant bias due to
such projection effects, even if it cannot be excluded that a
particular object might be affected.

According to predictions from numerical models, Jeans-fragmentation
should preserve the rotation of the underlying disk (see, e.g.,
\citealt{immeli04, bournaud07}). Therefore, it is expected that
distant clumpy galaxies should in principle appear as RD or PR (see,
e.g., \citealt{genzel08}). Indeed, clumps have diameters $\sim$kpc,
while the GIRAFFE IFU pixel size is 0.52 arcsec, which corresponds to
3.5 kpc at z=0.6. Therefore, the spatial scale of clumps remains well
below the spatial resolution of GIRAFFE observations. This means that
all kpc-sized substructures are strongly smoothed by the GIRAFFE IFU,
and it is therefore expected that most Jeans unstable disks show a
clear large-scale rotation, possibly affected by
perturbations\footnote{Simulations confirm this effect, as illustrated
  in the frame of the E-ELT by \cite{puech09d}, see their Fig. 13. See
  also \cite{bournaud08} and their Fig. 8: the smoothing applied to
  match observations is seven times smaller in length that the GIRAFFE
  IFU spatial resolution ($\sim$7 kpc), while all kinematic
  disturbances are already strongly attenuated at the kpc scale. Such
  a galaxy would clearly be classified as RD or PR, but not as CK
  because of the clear velocity gradient and the velocity dispersion
  peak located at the dynamical center.}. It cannot be excluded that
due to a given particular orientation or a particularly large number
of clumps, and due to the relatively coarse spatial resolution of the
GIRAFFE IFU, a given Jeans-unstable disk could not be classified as a
galaxy with a complex kinematics. However, this would require very
strong perturbations in the morpho-kinematics at a level which does
not appear to be representative of most clumpy disks observed at high
redshifts \citep{bournaud08}. Moreover, we find that 9\% of
intermediate-mass galaxies at z$\sim$0.6 are CK clumpy galaxies. If
clumps were strong enough to break a pre-existing disk, this would
mean that such galaxies would result in elliptical galaxies. Indeed,
given the low level of cold gas accretion at z$<$1 (see Sect. 4.2),
there would be simply not enough fresh material available to reform a
disk after its destruction. However, the fraction of elliptical
galaxies does not evolve significantly over this redshift range
\citep{hammer05,delgado10}. This implies that, if such strong clumps
exist, they cannot destroy completely the disk. Hence, a rotation
should be detected in such objects, particularly on large spatial
scales, such as those probed by the GIRAFFE IFU. To this respect,
there is a sub-class of CK galaxies which deserves more attention.
Those are the galaxies which show a velocity gradient that might be
associated with rotation, but which were classified as CK because of a
significant misalignment between their kinematic and morphological
main axes. In the sample of CK clumpy galaxies, only
J033224.60-274428.1 belongs to this sub-class (see Fig. \ref{kin}).
Such a particular case cannot affect significantly the conclusions of
this paper.

The above suggests that a very large majority of z$\sim$0.6 CK clumpy
galaxies are not the result of Jeans instability. Actually, such a
kinematics was rather associated with the result of major mergers, as
suggested by \cite{peirani09}, \cite{puech09}, \cite{hammer09b},
\cite{yang09}, and \cite{fuentes10}. During such processes, the
gaseous phase can be compressed to column densities high enough to
make the gas unstable, which can result in the formation of clump
complexes, as shown by \cite{elmegreen93}. Interestingly,
\cite{elmegreen93} mentioned that stronger perturbations are expected
to lead to a larger velocity dispersion and larger cloud complexes by
gravitational instability. If clumpy galaxies classified as CKs at
z$\sim$0.6 are the result of galaxy mergers, then one expects that
their mass ratios, orbits, and merger phases correspond to the
strongest perturbations, in terms of gas compression.

\cite{hammer09b} conducted a systematic first-order modelling of
z$\sim$0.6 galaxies in the IMAGES-CDFS sample. They used
hydrodynamical simulations from \cite{barnes02} to infer for each
galaxy a possible mass ratio, merger phase (i.e., before the first
pass, between the first and second pass, nuclei fusion, etc.), and
orbit. They used a limited base of reference simulations, but this is
enough to infer first-order conditions for the progenitors. Note that
in two cases, a full modelling was carried out by \cite{peirani09} and
\cite{fuentes10}, which confirms the first-order model inferred by
\cite{hammer09b}. For the five clumpy galaxies classified as CKs,
\cite{hammer09b} indeed found mass ratios between 1:4 and 1:1, merging
phases between the first passage and the nuclei fusion, and direct or
inclined orbits, i.e., the initial conditions that are expected to
result in the strongest gas compression during the collision, as
expected from numerical simulations of binary mergers (e.g.,
\citealt{dimatteo07,cox08}). One might however object that clumpy
galaxies with complex kinematics (and more generally z$\sim$0.6 CK
galaxies in the parent sample) do not necessarily show morphological
features like tails, bridges, double nuclei, nearby companions, etc.,
which are generally considered as clear evidences for mergers.
However, it is important to realize that available images, and even
those from the HST/ACS, are not deep enough to guarantee a systematic
detection of such features. For instance, with HST/ACS images in the
GOODS field, one would detect the optical radius of the MW only up to
z~0.5 (as inferred by redshifting R$_{25}$ with S/N=3 in a 1.5
resolution element of ACS). Hence, z$\sim$0.6 (i.e., 6 Gyr ago) is
probably a limit for detecting such low surface brightness features,
contrary to what is generally thought (see also \citealt{barden08} and
their Fig. 3). Therefore, the above-mentioned models can be considered
as reasonable evidence that clumps in galaxies showing complex
kinematics could be the result of gas compression induced by galaxy
mergers.

\subsection{Clumpy galaxies with rotation: Jeans-unstable disks?}
In this section, I investigate the stability of the six remaining
clumpy galaxies that were classified as RDs or PRs, i.e., those
showing large-scale rotation. The instability of gravitational disks
can be quantified using the Toomre parameter $Q$
\citep{safronov60,toomre64,goldreich65}. For a gravitational disk to
fragment into clumps, it is required that $Q<1$, which means that the
system is unstable to both radial and axisymmetric perturbations
(e.g., \citealt{polyachenko97,griv06}).

For a pure gaseous disk, and at large galacto-centric radius (i.e., on
the flat part of the rotation curve), $Q_{gas}$ can be directly
estimated from observables in the following way \citep{elmegreen93}:
$$ Q_{gas} = 1.4\frac{\left( \frac{\sigma _{gas}}{40 km.s^{-1}}\right)
  \left( \frac{V_{rot}}{200 km.s^{-1}} \right) }{\left(
  \frac{R_{gas}}{20 kpc}\right) \left(\frac{\Sigma _{gas}}{30M_\odot
    pc^{-2}}\right) } \,,
$$ where $\sigma _{gas}$ is the gas velocity dispersion, $\Sigma
_{gas}$ is the gas surface density, $V_{rot}$ is the gas rotation
velocity, and $R_{gas}$ is the gas extension. $V_{rot}$ were taken
from \cite{puech08}, and $R_{gas}$ from \cite{puech09b}. $\Sigma
_{gas}$ was estimated following \cite{puech09b}, i.e., by inverting
the Schmidt-Kennicutt law that relates the star formation rate and gas
surface densities. We refer the reader to \cite{puech09b} for a
complete description of the method. Finally, $\sigma _{gas}$ were
estimated following \cite{puech07}, i.e., as a sigma-clipped mean over
the velocity dispersion map, after having excluded the central pixel
which, at the spatial resolution of the GIRAFFE IFU, is strongly
contaminated by the contribution of larger-scale motions (see
\citealt{puech07} for details).

However, the use of the parameter $Q$ is complicated by the fact that
real galaxies are (at least at first order) a two-phase medium, and
cannot be considered as pure stellar or gaseous systems
\citep{jog84,jog84b}. Indeed, the median gas fraction in the IMAGES
sample is 31\% \citep{puech09b,hammer09b}. In such mixed systems, the
gravitational interaction between the stellar and the gaseous phase
makes the global system more unstable than the two phases considered
individually \citep{jog96}. In this case, the stability of such
systems cannot be derived from $Q_{gas}$ alone, and one has to define
an effective $Q_{eff}$ that takes into account the effect of both
stars and gas (e.g., \citealt{jog84,elmegreen95,jog96}). I used the
first-order approximation derived by \cite{wang94}:
$$
Q_{eff}^{-1} = Q_{gas}^{-1}+ Q_{stars}^{-1} \, ,
$$
with
$$
 Q_{stars} = 1.3\frac{\left( \frac{\sigma _{stars}}{40 km.s^{-1}}\right)
  \left( \frac{V_{rot}^{stars}}{200 km.s^{-1}} \right) }{\left(
  \frac{R_{stars}}{20 kpc}\right) \left(\frac{\Sigma _{stars}}{30M_\odot
    pc^{-2}}\right) } \, ,
$$ 
where the factor 1.3 in front of $Q_{stars}$ comes from the factor 1.4
in $Q_{gas}$ rescaled by $\pi$/3.36 (see, e.g.,
\citealt{elmegreen95}).
I converted half-light radii $R_{half}$ derived by \cite{neichel08}
from z-band HST/ACS images into total stellar radii $R_{stars}$ using
$R_{stars} = 1.9 \times R_{half}$, which corresponds to the optical
radius in thin exponential disks (see \citealt{puech09b} and
references therein). $\Sigma _{stars}$ was then derived using $\Sigma
_{stars} = M_{stellar} / \pi R_{stars}^2$, where $M_{stellar}$ are
stellar masses derived by \cite{ravi07} in the IMAGES sample (see also
\citealt{puech08,hammer09b}). I excluded contributions to stellar
masses from bulges by rescaling the stellar mass using the
bulge-to-total z-band light ratio B/T derived by \cite{neichel08}.
When no B/T could be measured, I assumed that all stars lie into a
disk. I assumed that $V_{rot}^{stars} \sim V_{gas}$, as generally
observed in local spiral galaxies \citep{vega01,pizzella04}. Finally,
in distant star-forming galaxies, it is very likely that
$\sigma_{gas}\sim \sigma _{stars}$, because a significant fraction of
stars were probably formed during the on-going starburst.

\begin{figure*}
\centering
\includegraphics[width=19cm]{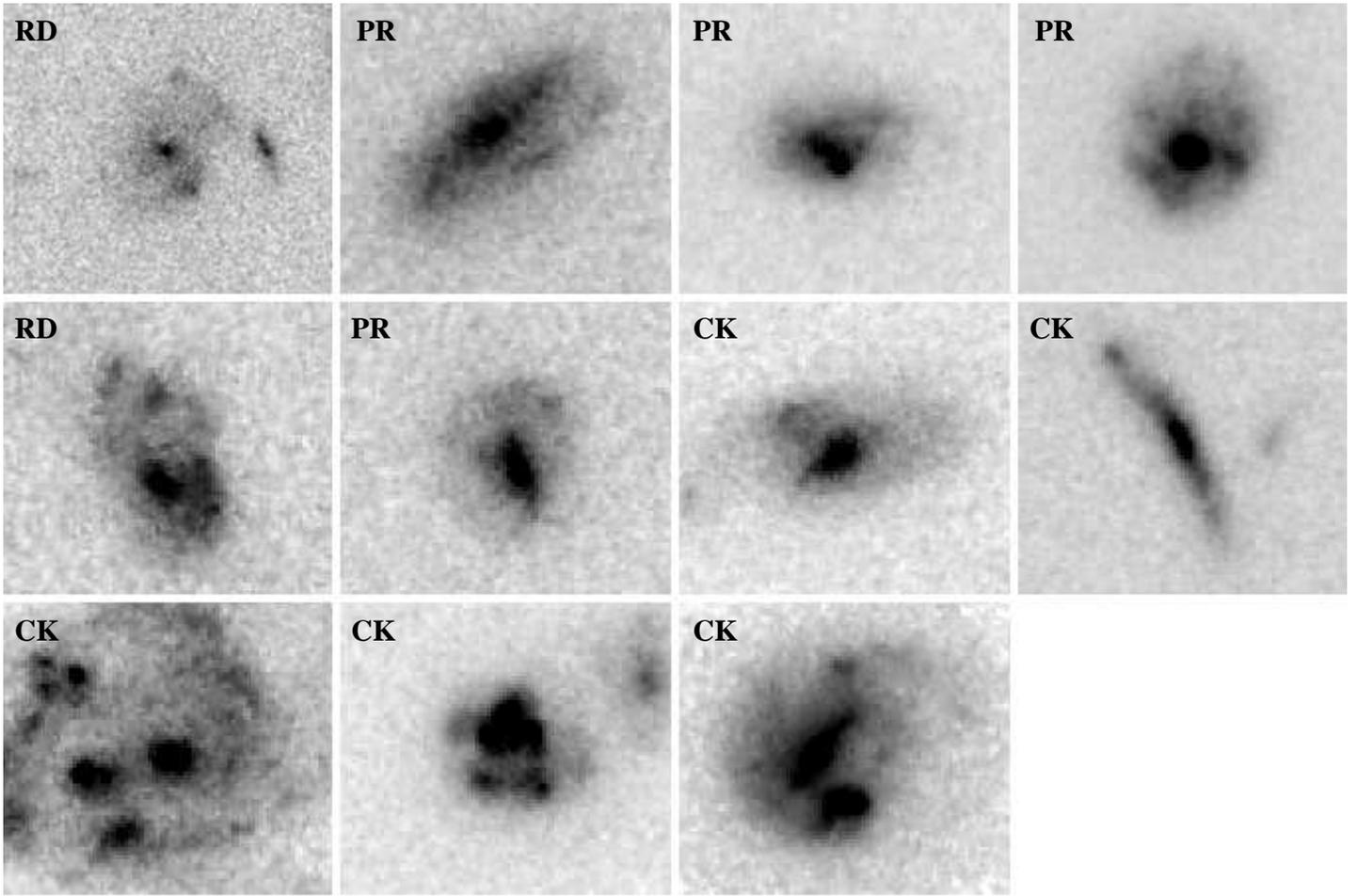}
\caption{Rest-frame optical ($\sim$5451\AA) morphology of the 11
  clumpy galaxies selected in the IMAGES survey, at z$\sim$0.6.
  \emph{From left to right, and top to bottom:} J033231.58-274121.6,
  J033210.25-274819.5, J033219.61-274831.0, J033233.90-274237.9,
  J033234.04-275009.7, J033239.04-274132.4, J033224.60-274428.1,
  J033227.07-274404.7, J033230.57-274518.2, J033234.12-273953.5,
  J033239.72-275154.7. The kinematic class of each object is indicated
  in the upper-left corner.}
\label{zband}
\end{figure*}

\begin{figure}
\centering
\includegraphics[width=9cm]{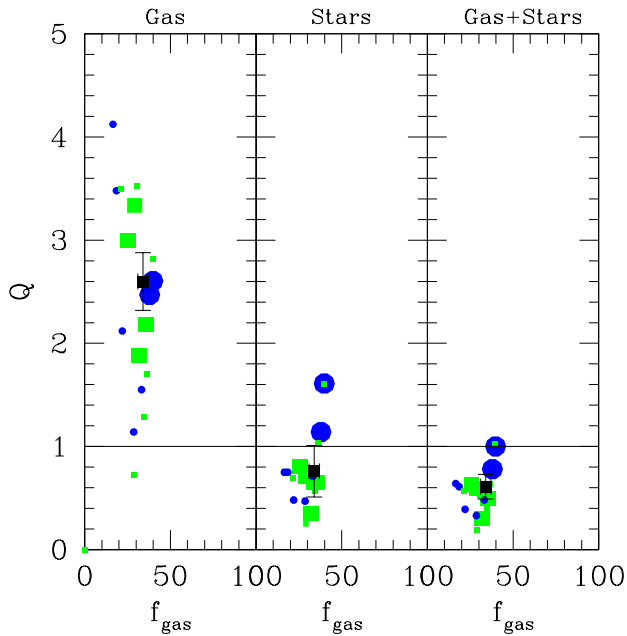}
\caption{Stability parameter $Q$ for the rotating disks (blue dots)
  and perturbed rotators (green squares) in the IMAGES-CDFS sample.
  The biggest symbols correspond to the six clumpy galaxies showing
  rotation; their median values are indicated by the black squares,
  with error-bars corresponding to 1-$\sigma$ bootstrap re-sampling.
  \emph{From left to right:} $Q_{gas}$, $Q_{stars}$, and $Q_{eff}$ as
  a function of f$_{gas}$. The marginal stability limit $Q$=1 is
  indicated as a solid black line.}
\label{stab}
\end{figure}

The resulting $Q_{gas}$, $Q_{stars}$, and $Q_{eff}$ are shown in Fig.
\ref{stab}. While the gaseous phase appears to be stable for the six
clumpy galaxies, the contribution of $Q_{stars}$ makes them globally
unstable to radial and azimuthal perturbations, i.e., to clump
formation. However, a global instability does not necessarily imply
that both the stellar and gaseous phases will fragment. If the
instability is not strong enough, only the gaseous phase (eventually
along with young hot O/B stars) can fragment, while the stellar
distribution (i.e., later type stars) remains much smoother. To check
this possibility, I examine z-band images, which corresponds to
rest-frame optical wavelengths (i.e., 5451\AA), for all the z$\sim$0.6
clumpy galaxies, as shown in Fig. \ref{zband}. There is a strong trend
in PR/RD clumpy galaxies to have their clumps vanishing in z band.
This is in contrast with higher-z clumpy galaxies, for which it has
been claimed that their morphologies do not change significantly at
longer wavelengths \citep{elmegreen07,lehnert09}. Note that clumpy
galaxies with complex kinematics tend to have persisting clumps in the
red. This is consistent with the fact that clumps in these galaxies
might result from merger-induced fragmentation, since, under this
hypothesis, they were found to correspond to the strongest conditions
in terms of gas compression (see above).

Interestingly, all RDs and PRs in the IMAGES-CDFS sample lie downward
the $Q_{eff}=1$ stability limit, while all do not harbor clumps, since
this includes those that were not selected as clumpy galaxies based on
their UV morphology (see Sect. 2). This could be linked with the
temporal evolution of instabilities in gas-rich disks. Indeed,
\cite{immeli04} studied the behavior of $Q_{gas}$, $Q_{star}$, and
$Q_{eff}$ in gas-pure unstable disks. Their simulations show that
Jeans-unstable disks are produced by early instabilities lasting a few
100 Myr, and driven by the gaseous phase (see their Fig. 3).
Conversely, instabilities first driven by the stellar phase tend to
result in stellar bars or spiral arms, depending on the radius where
the instability occurs. Therefore, at least some of the z$\sim$0.6
regular (i.e., not clumpy) RD or PR galaxies could be associated with
such initially star-driven instabilities, which will not result in
clumps but will rather develop bars or spiral arms. This could also be
due to a non-negligible contribution from the stellar mass by their
stellar halo, which can easily contribute to stabilize a disk
\citep{bournaud09}.

\section{Comparing clumpy galaxies at different redshifts}

\subsection{General properties of clumpy galaxies as a function of redshift}
In table \ref{comp}, I compare the main (median) properties of
z$\sim$0.6 and z$\sim$2 clumpy galaxies showing rotation. For
comparison at z$\sim$2, I adopted the sample of clumpy galaxies
studied by \cite{genzel08}, removing BzK6004-3482 and ZC782941 for
which no gas radius measurement were available. On overall, clumpy
galaxies at z$\sim$2 have a more massive and larger stellar phase than
rotating clumpy galaxies at z$\sim$0.6, which is reflected in their
larger $V_{rot}$, M$_{stellar}$, and R$_{half}$. They are also more
gas-rich, even relative to galaxies of similar stellar mass at
z$\sim$2 \citep{erb06}\footnote{\cite{erb06} derived gas masses
  considering a gas surface of FWHM$_{gas}^2$, while I used a surface
  of $\pi$r$_{gas}^2$. Assuming that the gas surface is Gaussian, then
  FWHM=2.35$\sigma$, while 4$\times \sigma$ encompasses 96\% of the
  total surface. I then find that, following \cite{erb06}, the median
  gas fraction in z$\sim$2 clumpy galaxies is 39\% instead of 65\% as
  quoted in Tab. \ref{comp}, which makes them gas-rich objects for a
  stellar mass of 10$^{11}$M$_\odot$ in comparison of their Fig. 16.
  However, our own-derived gas fractions appear to be in better
  agreement with the independent estimates of \cite{daddi09} in BzK
  galaxies at z$\sim$1.5 detected in CO.}. This larger gas content
make their gaseous phase Toomre-unstable ($Q_{gas}<$1), conversely to
rotating z$\sim$0.6 clumpy galaxies (see Sect. 3.2), while their
stellar phase is only marginally stable on average. This suggests that
z$\sim$2 clumpy galaxies got unstable in their gaseous phase first,
and then developed clumps as described by \cite{immeli04}. Because
their $Q_{eff}$ is on average a factor two lower than in z$\sim$0.6
clumpy galaxies, the fragmentation process is expected to be stronger
in z$\sim$2 clumpy galaxies. This could explain why z$\sim$2 clumpy
galaxies appear to be more clumpy at optical wavelengths, compared to
their z$\sim$0.6 counterparts (see Fig. \ref{zband} and above).

\begin{table}
\centering
\caption{Median properties of clumpy galaxies at z$\sim$0.6 (i.e.,
  those classified as PRs or RDs) and at z$\sim$2. \emph{From top to
    bottom:} rotation velocity, stellar mass (a diet Salpeter IMF was
  used by puech et al. (2008) at z$\sim$0.6, while a Chabrier IMF was
  used by Genzel et al. (2008); Stellar masses derived by Genzel et
  al. (2008) were converted into a diet Salpeter IMF using
  M$_{stellar}$(IMF$_{dietSalpeter}$)-0.093=M$_{stellar}$(IMF$_{Chabrier}$),
  following Gallazzi et al. (2008)), gas fraction (using the
  Schmidt-Kennicutt law of Kennicutt et al. (1989)), optical
  half-light radius (in z band for z$\sim$0.6 galaxies, and
  1.68$\times$R$_d$ for z$\sim$2 galaxies), gas radius (from Puech et
  al. 2009 at z$\sim$0.6 and Förster-Schreiber et al. 2006 at
  z$\sim$2), gas velocity dispersion (following Puech et al. 2007),
  and Toomre stability factor of the gas. }
\begin{tabular}{ccc}\hline
 & z$\sim$0.6 & z$\sim$2\\\hline
N$_{galaxies}$ & 6 & 6\\\hline
V$_{rot}$ [km/s] & 155 & 233\\
$\log{(M_{stellar})}$ [M$_\odot$] & 10.2 & 10.9\\
f$_{gas}$ [\%] & 34 & 65\\
R$_{half}$ [kpc] & 3.8 & 7.5\\
R$_{gas}$ [kpc] & 10.5 & 11.2\\
$\sigma _{gas}$ [km/s] & 39 & 69\\
Q$_{gas}$  & 2.6 & 0.5\\
Q$_{stars}$  & 0.8 & 1.0\\
Q$_{eff}$  & 0.6 & 0.3\\\hline
\end{tabular}
\label{comp}
\end{table}

Another difference between z$\sim$2 and z$\sim$0.6 clumpy galaxies is
that the former have a gas velocity dispersion twice as large as that
in the latter. The large gas velocity dispersions in z$\sim$2 clumpy
galaxies were interpreted as a combination of feedback from intense
star formation with turbulence generated by streams of cold gas as it
enters the forming disk \citep{genzel08,khochfar08}. However,
\cite{lehnert09} showed that cold flows do not have enough energy to
sustain the high gas velocity dispersion if gas dissipates energy as
compressible turbulence. Indeed, the only way for cold streams to
generate substantial dispersion in the gaseous phase is to contain
itself a clumpy component with gas densities similar to that of the
disk, and which dominates the stream in term of mass, but these do not
appear to allow a long-lived clumpy phase to take place in the disk
\citep{dekel09b}. Instead, the disk self-gravity itself could play an
important role in favoring the formation of massive clumps, as
suggested by \cite{burkert09}. However, \cite{lehnert09} estimated
that the disk self-gravity might indeed be powering strong turbulence
but only in an early stage. Such a turbulence might be driving the
collapse of massive clumps, while also triggering intense star
formation. \cite{lehnert09} argued that at this stage, the intense
star formation is likely self-regulated by the output of massive
stars, and this would take over self-gravity as the main driver for
the high velocity dispersions observed in z$\sim$2 clumpy galaxies
(see also \citealt{elmegreen09c}).

In comparison to z$\sim$2 galaxies, the pressure in the ISM of
z$\sim$0.6 clumpy galaxies appear to be much lower: the mean velocity
dispersion is twice lower, while the star formation density, with a
mean value of $\sim$0.02 M$_\odot$/yr/kpc$^{-2}$, is one order of
magnitude lower than at z$\sim$2 \citep{lehnert09}. Following
\cite{lehnert09}, one can roughly estimate that feedback from star
formation could account for 36-85\% of the median gas turbulence. For
five out of the eleven z$\sim$0.6 clumpy galaxies, it was possible to
compare the positions of emission and absorption lines using FORS2
spectra from \cite{rodrigues08}. Only J033224.60-274428.1 shows a
significant shift ($\sim$100km/s, \citealt{puech09b}), which suggests
substantial winds in this galaxy. Therefore, feedback from star
formation is probably not the main driver for the high velocity
dispersion observed in z$\sim$0.6 clumpy galaxies. Inter-clump gravity
could account for an additional 21-69\% contribution in z$\sim$0.6
clumpy galaxies \citep{lehnert09}, which is not sufficient to account
for the observed level of turbulence. Another source of turbulence
could be associated with cold gas accretion. In their semi-analytic
model, \cite{khochfar08} calibrated a relation between the cold gas
accretion rate and the associated generated gas velocity dispersion
using z$\sim$2 galaxies. Using the same calibration, I estimate that
to generate the observed level of gas turbulence in z$\sim$0.6 clumpy
galaxies, an accretion rate of $\sim$34M$_\odot$/yr would be required.
This is one order of magnitude above what is expected from full
numerical simulations (see \citealt{keres09} and discussion below in
Sect. 4.2). Finally, the turbulence in z$\sim$0.6 clumpy galaxies
could be merger-driven. Simulated gas-rich remnants from major mergers
(e.g., \citealt{robertson08}) indeed show similar $V_{rot}/\sigma$
ratios in the rebuilt disks compared to observed values
\citep{puech07}.

In order to relax towards local galaxies, the stellar phase of
z$\sim$0.6 clumpy galaxies needs to get stabilized by increasing
$Q_{stars}$ by at least $\sim$25\% on average. This could be achieved
by increasing $V_{rot}$ through gas accretion within the optical
radius. Alternatively, this can also be achieved through the
stabilizing influence of a growing bulge as described by
\cite{bournaud09}.
Finally, if
self-gravity drives disk turbulence \citep{burkert09}, then one
expects that z$\sim$0.6 clumpy galaxies get stabilized by increasing
their velocity dispersion to $\sim$49 km/s in order to reach
Q$_{eff}$=1. This is relatively close to the median value observed in
z$\sim$0.6 RD galaxies, with 47$\pm$3 km/s.

\subsection{Cold streams as a driver for clump formation}
Cold streams are thought to be an important process triggering
instability, as z$\sim$2 disks are expected to be fueled by such cold
streams, which could maintain a dense gaseous disk that can undergo
gravitational fragmentation into clumps \citep{dekel09b}. Indeed,
z$\sim$2 clumpy galaxies are thought to live in
$\sim$10$^{12}$M$_\odot$ dark matter haloes (e.g., \citealt{dekel09}),
in which such cold flows are expected to take place \citep{dekel09}.
To estimate the halo mass of the z$\sim$0.6 clumpy galaxies, we
assumed that rotation velocity is a good tracer of the circular
velocity of their halos, from which the halo mass can be estimated,
following the formalism of \cite{mo98}. Similar results were obtained
from halo abundance matching models \citep{conroy09,moster09}: we
indeed found that z$\sim$0.6 clumpy galaxies also reside in
$\sim$10$^{12}$M$_\odot$ haloes. Therefore, clumpy galaxies at
z$\sim$2 and z$\sim$0.6 inhabit halos of similar
mass\footnote{Simulations suggest that the local descendant of such
  z$\sim$0.6 halos are similar in mass, while the local descendant of
  z$\sim$2 clumpy galaxies halos are slightly more massive, with
  4$\times 10^{12}$M$_\odot$ \citep{mcbride09}. This could suggest
  that both populations of clumpy galaxies do not sample a common
  evolutionary path down to z=0, but rather correspond to two distinct
  local populations. While z$\sim$0.6 intermediate-mass galaxies were
  identified to be the progenitors of local early spirals
  \citep{hammer05}, most of z$\sim$2 clumpy galaxies halos could
  evolve toward local halos massive enough to host an elliptical
  galaxy, rather than a spiral disk (see also Sect. 4.1 of
  \citealt{genel08}).}. However, according to numerical simulations
\citep{keres05}, and theoretical expectations \citep{dekel09}, while
cold flows are expected to penetrate such halos at z$\sim$2, they are
expected to be strongly attenuated in haloes of similar mass at
z$\leq$1. The dominant mode of accretion in such z$\sim$0.6,
$\sim$10$^{12}$M$_\odot$ haloes is rather expected to be hot gas
accretion, since these haloes are too massive to allow spherical cold
gas accretion to be significant. The cosmological simulations of
\cite{keres09} indeed suggest that the average total accretion rate of
gas onto the central galaxies inhabiting $\sim$10$^{12}$M$_\odot$
halos dropped down by a factor $\sim$3 between z$\sim$2 and z$\sim$0.6
(see their Fig. 7), while the fraction of gas accreted cold versus hot
decreased by a factor $\sim$1.5 over this redshift range (their Fig.
8). As a consequence, according to these simulations, the cold gas
accretion rate typically dropped down from $\sim$7.5M$_\odot$/yr to
$\sim$1.5M$_\odot$/yr (see also \citealt{brooks09}). Moreover, the
simulations of \cite{keres09} suggest that mergers dominate the mass
growth of galaxies at z$<$1, which is in line with spatially-resolved
kinematics of z$\sim$0.6 galaxies (see Introduction). Finally, the
average velocity dispersion is found to be roughly the same between
rotating disks, perturbed rotators, and galaxies showing a complex
kinematics at z$\sim$0.6 \citep{puech07,epinat09b}, as well as in the
clumpy galaxies (see Tab. \ref{comp}): if cold flows were feeding
z$\sim$0.6 clumpy galaxies, then one would expect to detect an
increase of the velocity dispersion in these systems, which is not
observed. In summary, theoretical expectations are found to be
drastically different, when comparing z$\sim$2 and z$\sim$0.6 massive
haloes. While cold streams are expected to feed z$\sim$2 clumpy
galaxies in fresh gas and trigger the formation of clumps, this is
unlikely to be the dominant mechanism at z$\sim$0.6.

\subsection{Interactions as a driver for clump formation}
Numerical simulations have suggested that interactions could be also
an efficient trigger for instabilities in gaseous disks, resulting in
the formation of clumps \citep{dimatteo08}. Simulations revealed that
such a trigger is efficient only if the fragmenting disk in the
progenitor is already marginally stable before the interaction. The
interaction does not necessary result in a merger, but even in this
case, it is the distant interaction that triggers the instability, and
not the eventual subsequent merger \citep{dimatteo08}. Another
possibility is that distant clumpy galaxies with complex kinematics
could also be compact groups observed while they are merging
\citep{amram04,amram07}. Finally, disk rebuilding, i.e., the latest
phase of a gas-rich merger where the gas expelled during the process
falls back to reform a disk, could also be a good driver for clump
formation, especially in galaxies that do not show any obvious of
morphological or kinematic peculiarities (although the depth of images
might not be large enough, see Sect. 3.1). Note that in principle,
disk rebuilding does not necessarily require merging. Indeed, the disk
rebuilding phase involves mainly infalling gas, which can comes from
gas accretion from the intergalactic medium (e.g., through cold
streams, see below), or from the gas expelled during the merger
itself. However, given the low accretion rate at z$<$1, disk
rebuilding is probably systematically associated with merging at these
redshifts.

It is interesting to look for local counterparts of clumpy galaxies,
because of the much higher spatial resolution and sensitivity of
observations. As part of a morphological study of Markarian galaxies
\citep{markarian77}, \cite{casini76} identified ``a new class of
object, with UV emission, irregular clumpy structure, large dimension,
high luminosities and large internal motions'' compared to normal
irregular galaxies, which they called ``clumpy irregular'' (see also
\citealt{casini76b}). They suggested that these galaxies might be ``in
a turbulent or fragmented state, with large cells where the rate of
star formation is high''. Their description matches astonishing well
the definition of clumpy galaxies at high redshift (e.g.,
\citealt{elmegreen07}). A few other clump irregulars (cI) were later
identified by \cite{maehara88}. It is likely that other unidentified
cI exist, but they are clearly very rare locally, even among
interacting and peculiar galaxies \citep{casini76}: the number of
known cI so far is 11, among which only 7 are not known members of
clusters (Mrk 325, 7, 8, 432, 297, VV 523, and NGC 6120). Here we
consider only these latter 7 cI for comparison with higher-z galaxies
in the field. Interestingly, two of them (i.e., Mrk8 and 325) were
recently shown to be structurally similar to high-redshift clumpy
galaxies \citep{petty09}.

On overall, cI have stellar masses ranging between $10^{9.1}$M$_\odot$
and $10^{11.4}$M$_\odot$ with a median of $10^{10.2}$M$_\odot$,
according to their rotation velocities\footnote{as given by the
  hyperleda database.} and the local stellar-mass Tully-Fisher
relation \citep{puech08}. They are on overall gas-rich galaxies (e.g.,
\citealt{casini79,maehara88,garland07}), with HI gas fractions ranging
between 4 and 77\%, and a median of 40\%, i.e., a fraction usually
found in local dwarfs rather than in spiral galaxies
\citep{schombert01}. Those for which kinematic data are available show
a distorted rotation (Mrk 296, see \citealt{casini79}; Mrk 325, see
\citealt{garland07,PerezGallego09}; Mrk 297, see
\citealt{garland07,garcialorenzo08}; VV523, see \citealt{rampazzo05}).
These cI galaxies are almost always found to be merging or interacting
systems (Mrk 325, see
\citealt{duflot82,homeier99,conselice00,garland07}; Mrk 8, see
\citealt{casini76b,conselice00b}; Mrk 423, see \citealt{rothberg04};
Mrk 297, see
\citealt{taniguchi91,rothberg04,garland07,garcialorenzo08}; VV 523,
see \citealt{pustilnik03,rampazzo05}), or are part of a pair that
could explain their morpho-kinematic peculiarities (Mrk 7 and
NGC6120).

A similar class of objects was recently pointed up as presenting
similarities with distant clumpy galaxies. Those were selected as
super-compact UV Luminous galaxies by \cite{heckman05} and
\cite{hoopes07} in order to isolate Lyman Break Analogs (LBAs) of
distant Lyman Break Galaxies (LBGs). LBAs appear to be very rare
locally \citep{heckman05}, and indeed share many similarities with
distant LBGs, such as their (stellar or dynamical) masses, UV
luminosities, sizes, star-formation rates, dust attenuation, or gas
metallicities \citep{heckman05,hoopes07,basu07}. High-resolution
images from the HST revealed UV morphologies similar to those of LBGs,
once degraded to similar depth and spatial resolution
\citep{overzier08,overzier09b}. In the UV, they are also characterized
by complexes of massive clumps of star formation, while in the
optical, they appear to be dominated by features evidencing
post-mergers of interactions \citep{overzier09}. These low-surface
brightness material characteristic of mergers or interactions such as
faint companions or tidal features, are almost impossible to detect in
the redshifted images, which led \cite{overzier08} to suggest that
LBGs could be gas-rich mergers or interactions of relatively low-mass
galaxies. In a preliminary study of LBA kinematics, \cite{basu09}
found quite disturbed gas dynamics suggestive of feedback,
interaction, or merger events. They simulated how their kinematics
would appear at higher redshifts, and found similarities with
spatially-resolved kinematics of distant LBGs (e.g.,
\citealt{wright07,law09}).

LBAs have stellar masses ranging from $10^{9}$M$_\odot$ to
$10^{10.7}$M$_\odot$ with an average of $10^{10}$M$_\odot$ (see
\citealt{hoopes07,basu09}), i.e., similar to cI galaxies. Using the
relationship between SFR/M$_*$ (see \citealt{hoopes07} for typical
values in LBAs) and the gas-to-stellar mass ratio found in local
galaxies by \cite{zhang09}, one can infer gas fractions in LBAs, which
are found to be typically larger than 40-50\% , i.e., again in good
agreement with cI galaxies. Jeans instabilities in gas-rich disks
require relatively high gas densities to happen. As well as at
z$\sim$0.6 (see Sect. 4.2), cold streams are also suppressed in z=0
haloes more massive that $\sim$10$^{12}$M$_\odot$ (corresponding to
stellar masses larger than $\sim$10$^{10.5}$, as suggested by halo
occupation models; see, e.g., \citealt{conroy09}), but cold gas
accretion can still take place in a more spherical geometry in lower
mass haloes, such as those where cI and LBAs inhabit in on average.
This might be a mechanism by which dense gas disk might be maintained,
and fragmentation could follow. However, numerical simulations showed
that this mode of cold gas accretion is very low, which was confirmed
by observations, with $\sim$0.3M$_\odot$/yr \citep{sancisi08}. One
could therefore expect that local clumpy galaxies, just like their
z$\sim$0.6 counterparts, would be rather driven by interactions.
Interactions, and more generally collisions, are indeed the only way
to pressure the gas at level high enough to allow fragmentation,
except at very high redshifts, where cold streams can bring a lot of
cold gas directly to the disk within a dynamical time (see also
\citealt{elmegreen09b}).

\section{Discussion}
Spatially-resolved kinematics of distant galaxies suggested two
different mechanisms to account for galaxy evolution. While
observations at z$\sim$2 suggested that the dominant mechanism taking
place would be Jeans fragmentation leading to thick gas-rich, rotating
clumpy galaxies, observations at z$\sim$0.6 led to a different
conclusion, with interactions being the dominant driver, even in
clumpy galaxies, as suggested by the present study. Both populations
of clumpy galaxies are found to inhabit in halos with similar mass,
but numerical simulations and theoretical studies suggested that the
cosmological context evolved strongly during this redshift range:
while massive cold flows are expected to penetrate z$\sim$2 halos and
feed the central disks in cold gas, such cold streams are also
expected to be strongly suppressed in halos of same mass at z$\leq$1.
This explains why z$\sim$2 could indeed being undergoing a Jeans
instability phase, with cold flows maintaining a dense enough gaseous
disk resulting in the formation of clumps in a pre-existing rotating
disk. Such cold flows are almost suppressed at z$\sim$0.6 and
therefore cannot drive anymore the formation of clumps. Instead,
interactions, with or without a subsequent merger, can induce gas
compression high enough to result in the formation of clumps.
Detections of cold gas at high redshifts were claimed in a several
cases (e.g, \citealt{nilsson06,noterdaeme08,jorgenson09}), and recent
numerical simulations suggest that Ly-$\alpha$ blobs at high redshifts
might be a direct signature of cold streams \citep{goerdt09}. If the
theoretical frame and numerical simulations depicting cold flows turn
out to be correct, there would be no contradiction between the
interpretations of spatially-resolved kinematics of z$\sim$2 and
z$\sim$0.6 galaxies.

However, it is not clear whether mergers and/or interactions could
also account for the clumpy galaxies observed at z$\sim$2. Numerical
simulations showed how very gas-rich major mergers can also result in
similar star formation rates, gas surface density, and circular
velocity-to-velocity dispersion ratios that match observations
\citep{robertson08}. However, \cite{bournaud09} noticed that the
\cite{robertson08} simulation failed to reproduce the observed clumps
in these galaxies, which, they argue, can be attributed to the fact
that the remnant has too low a disk density to fragment. The numerical
simulation of \cite{robertson08} result in a bulge that indeed
contains 46\% of the stellar mass, which might explain why the rebuilt
disk did not fragment. \cite{bournaud09} further argued that mergers
systematically result in the build-up of a central stellar spheroid
that stabilizes the disk against fragmentation, which would rule out
merging as the underlying cause of fragmentation in z$\sim$2 clumpy
galaxies. However, \cite{bournaud09} models of unstable disks allow a
maximal fraction of 20\% and possibly up to 30\% of the stellar mass
to be in the bulge (and/or halo) for the fragmentation to occur.
Nevertheless, \cite{hopkins09b} showed that about 40-50\% of the
galaxies with stellar mass of 10$^{11}$M$_\odot$ at z$\sim$2 could
have a stellar bulge-to-disk ratio lower that 0.3, as a result of
mergers at higher redshift, which lets room for a significant fraction
of z$\sim$2 galaxies that could possibly result from major mergers
(see below). In z$\sim$2 galaxies, the median gas fraction is
estimated to be $\sim$50\% \citep{erb06,daddi09}, therefore, even
larger gas fraction \emph{in their progenitors} can be expected, which
will limit the effect of the violent relaxation in the remnant, hence
the bulge-to-disk ratio \citep{hopkins09b}. Moreover, interactions
between gas-rich, marginally unstable progenitors with similar masses
were shown to possibly result in clump formation \citep{dimatteo08}:
given the large gas fraction of z$\sim$2 galaxies, it is not
unreasonable to expect such a marginal stability in a large fraction
of them. It appears therefore possible that the z$\sim$2 clumpy
galaxies might result from distant interactions of gas-rich galaxies,
possibly followed by a merger.

In addition, while the fraction of clumpy galaxies at high redshift is
unambiguously large, it is still not clear whether they are all truly
rotating. \cite{forsterschreiber09} found that roughly one third of
galaxies in their sample are dominated by rotation\footnote{It must be
  mentioned here that their full sample is biased toward relatively
  massive star-forming galaxies. However, this fraction seems to hold
  when restricting their sample to stellar masses larger than
  2$\times$10$^{10}$M$_\odot$, for which the resulting sub-sample is
  truly representative of z$\sim$2 galaxies, see Fig. 4 of
  \cite{forsterschreiber09}.}, another third is found to be clear
interacting/merger systems, while the remaining third is found to
correspond to velocity-dispersion dominated systems. The latter tend
to show motions compatible with expectations from close mergers, at
least in a significant fraction of them \citep{forsterschreiber09}.
Strikingly, while their ``observed kinematics [...] are similar to
those of merger-driven starburst galaxies in the local universe'',
\cite{law09} claimed that these systems are probably resulting from
other processes probably related to gas accretion. The main argument
raised by \cite{law09} is that the major merger rate found in
cosmological simulations is not large enough to account for the
density of all star-forming galaxies at z$\sim$2. However, this
argument relies only on comparisons with semi-analytic simulations,
which are well know to be unable to reproduce a number of properties
of local galaxies (see, e.g., \citealt{dutton08}). Observations
suggest that the \emph{cumulative} fraction of merger that
M$_{stellar}>$10$^{10}$M$_\odot$ galaxies undergo between z=2 and z=3
(i.e., over 1.2 Gyr) is $\sim$2 \citep{conselice08}, meaning that
\emph{all} z$\sim$2 galaxies with M$_{stellar}>$10$^{10}$M$_\odot$,
i.e., as massive as clumpy galaxies, might potentially be the remnant
of a recent gas-rich major merger occurring at higher redshift. The
comparison between the spatially-resolved kinematics of local and
distant galaxies reveals that it remains difficult to assess
unambiguously the real nature of very distant systems
\citep{epinat09b}. The morphology and kinematics of local analogs of
some of these distant objects (i.e., the LBAs) reveal features
characteristic of mergers or interactions (see Sect. 4.3); once
projected at higher redshifts, these objects show striking
similarities with dispersion-dominated z$\sim$2 galaxies
\citep{overzier08,overzier09b,basu09}. This suggests that galaxy
interactions and/or mergers could also be driving the dynamics of
dispersion-dominated z$\sim$2 galaxies. This might potentially raise
the fraction of merger-driven galaxies to the two-thirds of the
z$\sim$2 galaxy population.

The different possible drivers for clump formation as a function of
redshift are summarized in Tab. \ref{sc}.

\begin{table}
\centering
\caption{Summary of the favored driver for clump formation at
  different redshifts. Two crosses denote a very good likelihood,
  while a single cross denotes a less unambiguous likelihood. A minus
  sign means a very low likelihood. The mechanism responsible for
  clump formation at z$\sim$2 is more uncertain than at lower redshift
  because of the larger uncertainty associated to the galaxy merger
  fraction at z$\sim$2, and of the lack of direct and unambiguous
  observational evidence of cold gas accretion in galaxy halos.}
\begin{tabular}{|c|c|c|c|}\hline
  & Interactions & Mergers & Cold gas accretion\\\hline
z$\sim$2 & + & + & +\\\hline
z$\sim$0.6 & ++ & ++ & -\\\hline
z=0 & ++ & ++ & -\\\hline
\end{tabular}
\label{sc}
\end{table}

\section{Conclusion}
I selected and studied 11 clumpy galaxies at z$\sim$0.6, drawn from
the representative IMAGES sample of emission line, intermediate mass
galaxies. Clumpy galaxies were selected mimicking the morphological UV
selection criteria used at larger redshifts, and their fraction at
z$\sim$0.6 was found to be $\sim$20\%. Among the 11 clumpy galaxies, 5
were found to show complex kinematics compatible with major mergers,
as suggested by \cite{hammer09b}. The remaining clumpy galaxies, i.e.,
which show large-scale rotation, are found to be Toomre-stable in
their gaseous phase, but unstable in their stellar phase, contrary to
higher-z clumpy galaxies, which are found to be unstable in both
phases, with a stronger level of effective instability. This could
originate both in the likely higher fraction of old stars in
z$\sim$0.6 clumpy galaxies, which would not participate into the
fragmentation process. This would naturally explain why z$\sim$0.6 UV
clumps tend to vanish when looking at reddest-band images, unlike
z$\sim$2 clumpy galaxies, which appear to be more persistent at longer
wavelengths.

While z$>$1 clumpy galaxies were largely associated with
Jeans-unstable disks in the literature, I argue that current kinematic
observations could actually support a fraction of only $\sim$33\% of
such systems among z$\sim$2 star-forming galaxies. While this is not
incompatible with the cumulative fraction of major mergers in such
objects over their past $\sim$1 Gyr, theoretical and numerical works
suggested a different channel for disk fragmentation, which could
result from cold flows from the inter-galactic medium penetrating
through the surrounding dark matter halos. These cold flows could
maintain the gaseous phase high enough to fragment and regenerate
clumps at a rate high enough to be consistent with observations. The
progressive shut down of these cold flows with redshift in massive
haloes could also explain why the formation of z$<$1 clumpy galaxies
are found to be preferentially driven by interactions, since the duty
cycle of merger would dramatically evolve between these two epochs.
Within this theoretical frame, there is therefore no contradiction
between the interpretation of kinematic observations of z$\sim$2 and
of z$\sim$0.6 galaxies. However, it remains unclear whether
interactions could not account for all or at least most of the
evolution seen in intermediate and massive galaxies all the way from
z$\sim$2-3 down to z=0.

Confirming the nature and the fraction of high-z clumpy galaxies will
require better spatial resolutions on larger telescopes.
Unfortunately, it is likely that such a debate will find a answer only
with the advent of the Extremely Large Telescopes and their associated
NIR integral field spectrographs such as EAGLE
\citep{puech08b,puech09c,puech09d}.

\section*{Acknowledgements}
This study made use of the \textsc{GOLDMINE} and \textsc{HYPERLEDA}
databases (see http://goldmine.mib.infn.it/ and
http://leda.univ-lyon1.fr/). I thank F. Bournaud for very valuable
comments on a preliminary version of this paper. I also thank F.
Hammer and H. Flores for their help in selecting clumpy galaxies in
the IMAGES survey, which initiated this work. I thank C. Balkowski
for a careful reading of this paper.

\appendix 

\section{Kinematical maps of z$\sim$0.6 clumpy galaxies}
The kinematical maps of the 11 clumpy galaxies discussed in Sect.
3 were published in \cite{yang08} as part of the IMAGES sample of
intermediate-mass, emission line galaxies at z$\sim$0.6. For the sake
of completeness, they are reproduced below.

\begin{figure*}
\centering
\includegraphics[angle=90,width=13cm]{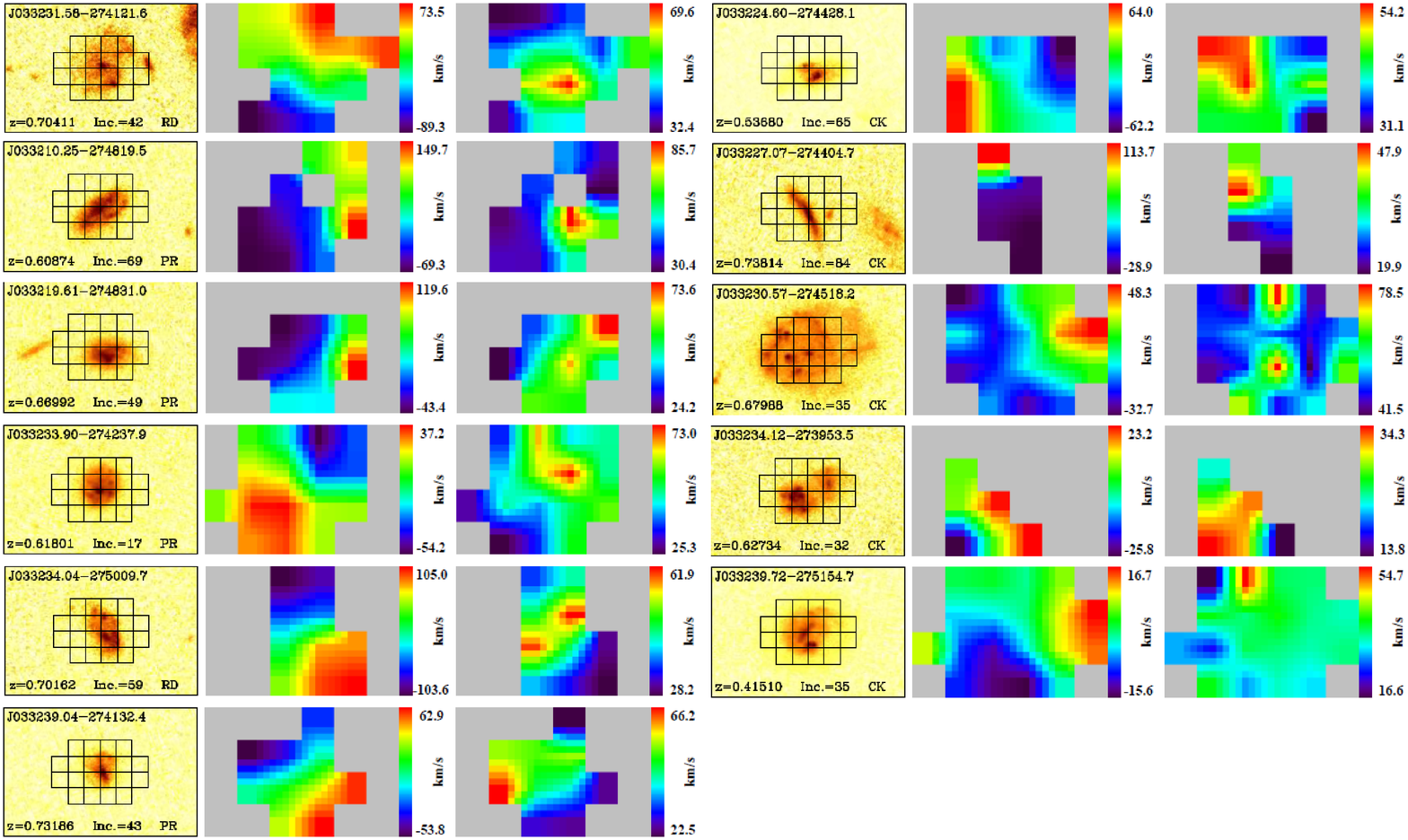}
\caption{Kinematics of the 11 clumpy galaxies observed by
  FLAMES/GIRAFFE at z$\sim$0.6 (reproduced from \citealt{yang08}).
  \emph{From left to right:}HST/ACS I-band images with the GIRAFFE IFU
  superimposed (0.52 arcsec/pix), velocity field (with a 5$\times$5
  linear interpolation), and velocity dispersion map. The first column
  corresponds to the 5 clumpy galaxies showing rotation (RD/PR), while
  the second column corresponds to CK galaxies (see Sect. 3).
  J033239.72-275154.7, and J033227.07-274404.7 were studied and
  modelled in detail by Peirani et al. (2009) and Fuentes-Carrera et
  al. (2010), respectively.}
\label{kin}
\end{figure*}

\label{lastpage}

\end{document}